# AI, insurance, discrimination and unfair differentiation: an overview and research agenda

Marvin van Bekkum, Frederik Zuiderveen Borgesius & Tom Heskes

# AI, insurance, discrimination and unfair differentiation: an overview and research agenda

Marvin van Bekkum, Frederik Zuiderveen Borgesius and Tom Heskes

Interdisciplinary Hub for Digitalization & Society (iHub) & Institute for Computing and Information Sciences (iCIS), Radboud University, Nijmegen, the Netherlands

**ABSTRACT**

Insurers underwrite risks: they calculate risks and decide on the insurance price. Insurers seem captivated by two trends enabled by Artificial Intelligence (AI). First, insurers could use AI for analysing more and new types of data to assess risks more precisely: data-intensive underwriting. Second, insurers could use AI to monitor the behaviour of individual consumers in real-time: behaviour-based insurance. For example, some car insurers offer a discount if the consumer agrees to being tracked by the insurer and drives safely. While the two trends bring many advantages, they may also have discriminatory effects on society. This paper focuses on the following question. Which effects related to discrimination and unfair differentiation may occur if insurers use data-intensive underwriting and behaviour-based insurance?

## 1. Introduction

Insurers offer important services to modern societies. For example, motor insurance is important for people's mobility, household insurance for protecting property, and life insurance for protecting family members against poverty.[1] Meanwhile, most insurers are companies aiming for profit. Consumers pay a price to an insurer to have their risks covered. Based on, for example, the consumer's car age and weight, insurers calculate the risk

**CONTACT** Marvin van Bekkum marvin.vanbekkum@ru.nl Radboud University, Erasmuslaan 1, 6525 GE Nijmegen NL, the Netherlands

[1]European Insurance and Occupational Pensions Authority (EIOPA), *Artificial Intelligence Governance Principles, Towards Ethical and Trustworthy Artificial Intelligence in the European Insurance Sector: A Report from EIOPA's Consultative Expert Group on Digital Ethics in Insurance* (Publications Office 2021) 24 <https://data.europa.eu/doi/10.2854/49874>.

that the consumer will file a claim, and base the premium (the price of insurance) on the risk.[2]

We identify two AI-related trends in the insurance sector. (i) The first trend is data-intensive underwriting: insurers experiment with AI to calculate risks and to set insurance prices. For example, insurers could analyse new types of data with the help of AI. (ii) The second trend is behaviour-based insurance: insurers increasingly monitor the behaviour of individual consumers.[3] In this paper, we explore the following question: Which effects related to discrimination and unfair differentiation may occur if insurers use data-intensive underwriting and behaviour-based insurance?

Our main contributions to the literature are as follows. First, as far as we know, we are the first to compare both trends in detail.[4] Our paper makes the point that many threats concerning AI systems are not necessarily about machine learning as a specific technology, but rather about organisations using vast datasets to make decisions about people. Second, in our analysis, we include insights from legal, philosophical, sociological and computer science literature.[5] We aim to avoid unnecessary jargon in our writing, making the paper more accessible to non-specialists.

Third, we analyse both trends, discussing them from the perspectives of non-discrimination law and fairness.[6] We assess the fairness of the two trends largely based on existing literature about insurance fairness. We use the distinction between discrimination (in the narrow legal sense) and unfair differentiation as we proposed in earlier work.[7] Finally, we provide a research agenda.

[2]In the rest of the paper, we use the word *price* instead of *premium*. Insurance Europe, *How Insurance Works (Booklet)* (Insurance Europe, 2012) <https://insuranceeurope.eu/publications/729/how-insurance-works/>.

[3]For example, some life and health insurers give discounts to consumers wearing fitness trackers. See e.g. Maria Henkel, Tamara Heck and Julia Göretz, 'Rewarding Fitness Tracking – The Communication and Promotion of Health Insurers' Bonus Programs and the Use of Self-Tracking Data' in Gabriele Meiselwitz (ed), *Social Computing and Social Media. Technologies and Analytics*, vol 10914 (Springer International Publishing, 2018) <https://link.springer.com/10.1007/978-3-319-91485-5_3>. Steffen Krüger and Niahmh Ní Bhroin, 'Vital Signs: Innovations in Self-Tracking Health Insurance and Social Change' (2020) 6 *The Journal of Media Innovations* 93 <https://journals.uio.no/TJMI/article/view/7836>. Andrea Martani, David Shaw and Bernice Simone Elger, 'Stay Fit or Get Bit – Ethical Issues in Sharing Health Data with Insurers' Apps' [2019] *Swiss Medical Weekly* <https://smw.ch/index.php/smw/article/view/2641>.

[4]Most other papers consider one of the two trends, or do not contrast the trends. For examples in other work, see e.g. Marta Infantino, 'Big Data Analytics, Insurtech and Consumer Contracts: A European Appraisal' (2022) 4 *European Review of Private Law* 613. Alberto Cevolini and Elena Esposito, 'From Pool to Profile: Social Consequences of Algorithmic Prediction in Insurance' (2020) 7 *Big Data & Society* 205395172093922 <https://journals.sagepub.com/doi/10.1177/2053951720939228>. Anya ER Prince and Daniel Schwarcz, 'Proxy Discrimination in the Age of Artificial Intelligence and Big Data' (2019) 105 *Iowa Law Review* 62. Arthur Charpentier, *Insurance: Discrimination, Biases & Fairness* (Institut Louis Bachelier, 2022) <www.institutlouisbachelier.org/en/pdf-reader/?pid=81252>.

[5]The paper defines AI based on technical definitions and discusses causality in the paragraphs concerning 'seemingly irrelevant characteristics'. For a more technical discussion, see Sander Greenland, 'Causality Theory for Policy Uses of Epidemiological Measures' in AD Lopez and Colin D Mathers, *Summary Measures of Population Health* (World Health Organization, 2002). Our paper also mentions the fair machine learning literature where relevant, such as in Section 4.2.2.

[6]Due to length limitations, we will not apply all possible notions of fairness in insurance. For example, we do not explicitly discuss 'solidarity'. We apply certain elements of solidarity, such as exclusion.

A few remarks about the scope of the paper. Where we write 'insurer', we mean any type of company offering insurance. This could include a tech startup just entering the insurance business, but also a centuries-old insurance company. Our paper focuses on insurance underwriting, the insurance industry's core business. Since we focus on discrimination and other unfair differentiation, many questions are outside the scope of the paper. For example, we do not discuss privacy or data protection, and we do not discuss the right to healthcare (relevant for health insurance).[8] We do not discuss all possible AI applications. For example, AI-driven fraud detection is outside the scope of the paper. We focus mostly on countries within the European Union. However, the paper could be relevant outside Europe.[9] We introduce some aspects of non-discrimination law, but do not discuss other fields of law. The paper focuses on AI and the processing of vast amounts of data in the insurance sector, but can also be useful for research and policy in other sectors working with AI: the insurance sector has decades of experience with statistics, AI and the regulation of AI.[10] Some scholars see insurance as 'an insightful analogue for the social situatedness and impact of machine learning systems'.[11]

The paper is structured as follows. We first summarise (in Section 2) how insurance functions and discuss data-intensive underwriting and behaviour-based insurance. In Section 3, we discuss the difference between non-discrimination and fairness. In Sections 4 and 5, we discuss non-discrimination and fairness aspects of the two trends. In section 6, we present a research agenda. Section 7 concludes.

[7] F Zuiderveen Borgesius, *Discrimination, Artificial Intelligence, and Algorithmic Decision-Making*. Report for the European Commission against Racism and Intolerance (ECRI) (Council of Europe, 2019) 7, 67 <www.coe.int/en/web/artificial-intelligence/-/news-of-the-european-commission-against-racism-and-intolerance-ecri->.

[8] See about the GDPR, data privacy and insurance e.g. Florent Thouvenin and others, 'Big Data in the Insurance Industry: Leeway and Limits for Individualising Insurance Contracts' <www.zora.uzh.ch/id/eprint/179171>; Nele Stroobants and Caroline Van Schoubroeck, 'Telematics Insurance: Legal Concerns and Challenges in the EU Insurance Market' (2021) 13 *European Journal of Commercial Contract Law* 51 <www.ingentaconnect.com/content/10.7590/187714621X16463138640038>; Jon Truby, Rafael Dean Brown and Imad Antoine Ibrahim, 'Regulatory Options for Vehicle Telematics Devices: Balancing Driver Safety, Data Privacy and Data Security' (2023) *International Review of Law, Computers & Technology* 1 <www.tandfonline.com/doi/full/10.1080/13600869.2023.2242671>. Z Bednarz, K Lewis and J Sadowski, '"It's Not Personal, It's Strictly Business": Behavioural Insurance and the Impacts of Non-Personal Data on Individuals, Groups and Societies' (2025) 56 Computer Law & Security Review 106096 <https://www.sciencedirect.com/science/article/pii/S0267364924001614>.

[9] For example, in the United States, some insurers collaborate with car manufacturers to include behaviour-based insurance in smart cars. See <www.nytimes.com/2024/03/11/technology/carmakers-driver-tracking-insurance.html>.

[10] Oscar H Gandy, *Coming to Terms with Chance: Engaging Rational Discrimination and Cumulative Disadvantage* (Routledge, 2016). Dan Bouk, *How Our Days Became Numbered* (University of Chicago Press, 2015).

[11] See Christian Fröhlich and Robert C Williamson, 'Insights From Insurance for Fair Machine Learning', The 2024 ACM Conference on Fairness, Accountability, and Transparency (ACM 2024) s 7 <https://dl.acm.org/doi/10.1145/3630106.3658914>.

## 2. Insurance, AI, and two trends

This section discusses insurance underwriting (2.1), AI for insurance underwriting (2.2), two trends related to AI in insurance (2.3 and 2.4) and the similarities and differences between the two trends (2.5).

### *2.1. Underwriting*

The core business of insurers is underwriting risks: estimating the expected claims cost of a consumer via a risk assessment.[12] Roughly summarised, risk-based underwriting works as follows. First, the insurer collects characteristics about the individual consumer, such as the consumer's car type for car insurance. The insurer assigns the consumer to a *risk pool*, a group of consumers with the same characteristics. The insurer estimates the probability that the consumer might file a future claim and the amount in Euros of the claim: the expected claim cost. The insurer estimates the claim cost on, for example, past data about the severity (size) of claims and the number of claims within the same risk pool.[13]

To illustrate with a simple example, take a consumer Alice who has one characteristic: 'drives in a light car'. A car insurer assigns a low estimated claim cost to Alice, because (as shown by past data) consumers with a light car cause less damage in an accident. The insurer charges Alice a low price (plus possible extra costs or discounts). Let us now take the same example for another consumer, Bob. Bob has two characteristics: 'drives in a light car' and 'is senior'. The underwriter assumes a higher estimated claim cost for Bob than for Alice, because (as shown by past data) older drivers are more often involved in accidents. The insurer decides that Bob's risk pool represents a medium risk. The insurer charges Bob a medium price (again, plus possible extra costs or discounts).[14]

[12]The Geneva Association, the global association of insurance companies, describes underwriting as 'a core process of insurance that involves assessing and pricing risks presented by applicants seeking insurance coverage'. The Geneva Association, *Research Brief. Promoting Responsible Artificial Intelligence in Insurance* (The Geneva Association – International Association for the Study of Insurance Economics) 17 <www.genevaassociation.org/sites/default/files/ai_in_insurance_web_0.pdf>.

[13]Insurance Europe, *Risk-Based Underwriting. How the Losses of the Few Are Spread among the Many* (2021) <www.insuranceeurope.eu/priorities/2473/risk-based-underwriting-incl-ec-beating-cancer-plan>. Insurance Europe, *Risk Pooling Information Sheet* (2021) <www.insuranceeurope.eu/downloads/risk-based-underwriting-pooling/Pooling.pdf>. Insurance Europe (n 2) 12. Some insurers also assign a weight to the rating factors. EIOPA, *Big Data Analytics in Motor and Health Insurance: A Thematic Review* (Publications Office of the European Union, 2019) 34.

[14]In practice, the risk could be more granular than only 'low, medium or high', such as a number between 0 and 1.

### 2.2. AI systems in insurance

Many insurers use statistical regression models to analyse datasets.[15] The insurer fits such a model to the data: for different consumer characteristics, the model should reflect the claims in the dataset as accurately as possible.[16] For example, a car insurance model could express that elderly drivers represent a higher risk, which follows from the data. We will use the term 'traditional' for such underwriting. Another example of traditional underwriting could be an insurer who finds a statistical correlation between postal codes and the number or burglaries. The insurer can then use the postal code for risk pooling in, e.g. home insurance.

Many insurers also use forms of AI to underwrite risks or to analyse past data. AI can be defined as 'the science and engineering of making intelligent machines, especially intelligent computer programs. […] Intelligence is the computational part of the ability to achieve goals in the world'.[17] According to the Geneva Association, the international association of insurance companies, a promising type of AI for underwriting is machine learning.[18] The promise of such systems is that insurers can more accurately analyse more data. Machine learning systems can 'learn from data and automate the process of analytical model building and solve associated tasks'.[19] Machine learning models could identify new attributes that indicate risk in datasets, without the insurer defining in advance which attributes the machine learning system should consider.[20] The insurer could calculate prices based on the new attributes. The underwriting process itself remains largely the same: insurers charge higher prices to riskier consumers.[21]

In countries such as the United Kingdom, insurers increasingly use machine learning for their underwriting practices.[22] According to the UK

[15] Autoriteit Financiële markten (AFM) and De Nederlandse Bank (DNB), Artificial Intelligence in the Insurance Sector. An Exploratory Study (AFM & DNB 2019) 8 <www.afm.nl/~/profmedia/files/rapporten/2019/afm-dnb-verzekeringssector-ai-eng.pdf>.

[16] A model is a representation of reality.

[17] John McCarthy, 'What Is AI? Basic Questions' <http://jmc.stanford.edu/artificial-intelligence/what-is-ai/index.html> accessed 30 January 2025.

[18] The Geneva Association (Noordhoek), *Regulation of Artificial Intelligence in Insurance: Balancing Consumer Protection and Innovation* (The Geneva Association, 2023) 10, 15, 17–19 <www.genevaassociation.org/publication/public-policy-and-regulation/regulation-artificial-intelligence-insurance-balancing>.

[19] Christian Janiesch, Patrick Zschech and Kai Heinrich, 'Machine Learning and Deep Learning' (2021) 31 *Electronic Markets* 685, 685 <https://link.springer.com/10.1007/s12525-021-00475-2>.

[20] The Geneva Association (Noordhoek) (n 18) 10–11.

[21] See Section 2.1. See also The Geneva Association (Noordhoek) (n 18) 17. We note that insurers could also automate other processes that affect the acceptance of consumers (non-risk-based pricing). For example, an insurer could use an (machine learning) algorithm to pre-approve customers based on their bank account data. Or an insurer could assess the willingness to pay of the consumer. See James Davey, 'Insurance and Price Regulation in the Digital Era' in TT Arvind and Jenny Steele (eds), *Contract Law and the Legislature: Autonomy, Expectations, and the Making of Legal Doctrine* (Hart Publishing, 2020) <https://eprints.soton.ac.uk/433153/>.

[22] David Piesse, *Embedded Artificial Intelligence (AI) in Financial Services* (International Insurance Society, 2023) <www.internationalinsurance.org/insights_cyber_embedded_artificial_intelligence_in_financial_services>. See on a global scale Rodrigo M Jesus, Miguel A Brito and Duarte N Duarte, 'Machine

Financial Conduct Authority, a UK insurer uses machine learning-based application to pre-approve consumers for life insurance using data already available from bank accounts and credit ratings.[23] In other countries, such as the Netherlands, insurers seem more cautious of using machine learning systems for underwriting.[24] In the following sections, we describe two AI-related trends in the insurance sector.

### 2.3. Trend 1: data-intensive underwriting

The first AI-related trend is data-intensive underwriting: insurers use and analyse more data for estimating the odds that a consumer files a claim and calculating the price based on that estimation. Insurers could use AI for underwriting in two different ways. First, insurers could use AI systems to analyse the data they already have. Second, insurers could use AI systems to analyse more data and new types of data. For more than a century, insurers gathered data directly from the consumers: for example, demographic data such as a person's age, or the consumer's smoking and drinking habits.[25] Insurers used questionnaires to gather consumer characteristics such as the type of car they drive.[26]

Now, insurers could analyse more data gathered from many different sources. For example, insurers could analyse a consumer's web searches on the internet (online media data), a consumer's shopping habits (bank account data), or selfies to estimate the consumer's age.[27] To illustrate, Dutch insurers expect to use social media data, bank account data, and data gathered from the Internet of Things devices.[28] Furthermore, insurers could analyse data with machine learning to predict risks more accurately.[29]

Many insurers have high hopes for data-intensive underwriting, but the trend is only slowly catching on in Europe. The European Union's data protection laws may slow down the abilities of insurers to gather the new types

Learning on Insurance Price Prediction', *Proceedings of the 2023 9th International Conference on Computer Technology Applications* (ACM 2023) <https://dl.acm.org/doi/10.1145/3605423.3605450>.

[23]Bank of England & Financial Conduct Authority (FCA), *Machine Learning in UK Financial Services* (2022) s 5.3 <www.bankofengland.co.uk/report/2022/machine-learning-in-uk-financial-services>.

[24]Autoriteit Financiële markten (AFM) and De Nederlandse Bank (DNB) (n 15) 8.

[25]EIOPA (n 13) 9 and further.

[26]Laurence Barry and Arthur Charpentier, 'Personalization as a Promise: Can Big Data Change the Practice of Insurance?' (2020) 7 *Big Data & Society* 205395172093514, 3–6 <https://journals.sagepub.com/doi/10.1177/2053951720935143>.

[27]EIOPA (n 13) 9 and further. More examples are possible, even emotion detection: Andrew McStay and Lachlan Urquhart, 'In Cars (Are We Really Safest of All?): Interior Sensing and Emotional Opacity' (2022) 36 *International Review of Law, Computers & Technology* 470 <www.tandfonline.com/doi/full/10.1080/13600869.2021.2009181>.

[28]Autoriteit Financiële markten (AFM) and De Nederlandse Bank (DNB) (n 15).

[29]See Section 2.1. EIOPA (n 13) 12.

of data compared to other countries.[30] And there seem to be practical limitations. For instance, Dutch insurers are not yet convinced of the value of social media data.[31]

However, the limitations will not prevent data-intensive underwriting altogether. If insurers find new insights from applying machine learning to large datasets, we can expect more data-intensive underwriting in Europe as well. The Dutch financial authority AFM expects the trend to catch on in the Netherlands and the rest of Europe:

> Digitisation is undeniably going to have an impact on the insurance sector and its supervision. Insurers are gathering more data, applying smarter data analytics, and in doing so influence their entire operations and distribution, from pricing to claims handling. This applies not only to Dutch insurers but to many European parties […].[32]

In sum, the first insurance trend enabled by AI is data-intensive underwriting.

### 2.4. Trend 2: behaviour-based insurance

The second AI-related trend is that insurers adapt the price based on how the consumer behaves in real-time: behaviour-based insurance. For example, a life or health insurer may offer a discount to consumers who, according to their health tracker, walk a certain number of steps.[33] A car insurer may offer a discount to consumers who, according to a device in their car, drive safely.[34] Such car insurances are often called telematics insurance.

These two examples of behaviour-based insurance are a first step towards individualising risk. Scholars started writing about such individualisation since the 80's.[35] In 1981, Walters stated that 'since the insurer assumes the *individual* insured's risk of loss, the price should be fundamentally based upon the expected value of an insured's losses'.[36]

[30] Infantino (n 4); Thouvenin and others (n 8) 242.
[31] Autoriteit Financiële markten (AFM) and De Nederlandse Bank (DNB) (n 15).
[32] Translated by authors. Autoriteit Financiële markten (AFM), *Technologie richting 2033. De toekomst van verzekeren en toezicht* [Technology Towards 2033. The Future of Insurance and Supervision] (AFM, 2023) 3 <www.afm.nl/nl-nl/sector/actueel/2023/april/kansen-risico-digitalisering-verzekeringsmarkt>.
[33] Gert Meyers, *Behaviour-Based Personalisation in Health Insurance: A Sociology of a Not-Yet Market* (KU Leuven, 2018) 18.
[34] The device measures, for instance, how often someone brakes hard, how fast someone drives, and how someone steers. Alberto Cevolini and Elena Esposito, 'From Actuarial to Behavioural Valuation. The Impact of Telematics on Motor Insurance' (2022) 9 *Valuation Studies* 109, 111–12 <https://valuationstudies.liu.se/article/view/3697>. Gert Meyers and Ine Van Hoyweghen, '"Happy Failures": Experimentation with Behaviour-Based Personalisation in Car Insurance' (2020) 7 *Big Data & Society* 205395172091465, 4 <https://journals.sagepub.com/doi/10.1177/2053951720914650>.
[35] See Barry and Charpentier, 'Personalization as a Promise' (n 26) 5; Sylvestre Frezal and Laurence Barry, 'Fairness in Uncertainty: Some Limits and Misinterpretations of Actuarial Fairness' (2020) 167 *Journal of Business Ethics* 127, 129 <https://link.springer.com/10.1007/s10551-019-04171-2>; Michael A Walters, *Risk Classification Standards* (Proceedings of the Casualty Actuarial Society, 1981) <www.casact.org/abstract/risk-classification-standards>.
[36] Walters (n 35) 3.

In recent years, new behaviour-based insurance products are being introduced. European insurers are interested in personalising the insurance of their consumers, and reports by insurers highlight behaviour-based insurance as an important trend.[37] Insurance could become more and more behaviour-based in the future: a 'not-yet market'.[38]

The prevalence of behaviour-based insurance varies from country to country in Europe. Italy was an early adopter of behaviour-based insurance. Italian insurance companies successfully introduced behaviour-based insurance by offering extra services and discounts to the originally high price of the Italian motor third-party liability insurance.[39] Another early adopter is the United Kingdom. The UK Financial Conduct Authority writes: 'Some [insurers] use telematics risk modelling based on Deep Neural Networks to estimate driver behaviour and, thereby, predict the magnitude of the claim and determine the prices charged to the consumer.'[40] In the Netherlands, behaviour-based car insurance is rarer.[41]

So far, in Europe, insurers have not adopted behaviour-based insurance for all their (car or life) insurance. One factor is cost: it costs too much for insurers to keep track of the behaviour of individual consumers, and creating a behaviour-based infrastructure also comes with high fixed costs for collecting data and administration.[42] While in this section we focus on the existing products in behaviour-based car and life insurance, insurers could use other Internet of Things (IoT) devices to offer more behaviour-based insurance in the future.[43]

The distinction between data-intensive underwriting and behaviour-based insurance is not clear-cut. For example, insurers could combine the two trends.[44] One can think of borderline cases, with characteristics of both practices. For example, insurers can combine real-time behavioural analysis with further data analysis of the driving habits of young drivers over a certain period.[45]

[37]The Geneva Association (Noordhoek) (n 18).

[38]Meyers (n 33).

[39]Swiss Re, *Unveiling the Full Potential of Telematics – How Connected Insurance Brings Value to Insurers and Consumers: An Italian Case Study* (Swiss Re, 2017) 7 <www.researchgate.net/publication/329775370>.

[40]Bank of England & Financial Conduct Authority (FCA) (n 23) s 5.3.

[41]A rare example of a behaviour-based insurance product in The Netherlands is <www.anwb.nl/verzekeringen/autoverzekering/veilig-rijden>.

[42]Infantino (n 4) 623.

[43]In 2014, Tim O'Reilly, founder and CEO of O'Reilly Media, said: 'I think that insurance is going to be the native business model for the Internet of Things'. Rik Myslewski, 'The Internet of Things Helps Insurance Firms Reward, Punish' <www.theregister.com/2014/05/23/the_internet_of_things_helps_insurance_firms_reward_punish/> accessed 28 January 2025.

[44]In her pop-sci book, O'Neil warns for insurers' feeding behavioral data into AI systems. See Cathy O'Neil, *Weapons of Math Destruction: How Big Data Increases Inequality and Threatens Democracy* (Penguin Books, 2016) 139.

[45]Roel Henckaerts and Katrien Antonio, 'The Added Value of Dynamically Updating Motor Insurance Prices with Telematics Collected Driving Behavior Data' (2022) 105 *Insurance: Mathematics and Economics* 79 <https://linkinghub.elsevier.com/retrieve/pii/S0167668722000385>. Behaviour combined with contextual information, see Sobhan Moosavi and Rajiv Ramnath, 'Context-Aware Driver Risk Prediction with Telematics Data' (2023) 192 *Accident Analysis & Prevention* 107269 <https://linkinghub.

Nevertheless, because of their different effects on the pricing of insurance, we discuss the two trends separately in the paper.

### 2.5. The two trends: similarities and differences

Data-intensive underwriting and behaviour-based insurance are both made possible, in part, by developments in AI. We highlight two differences between traditional insurance and the two trends.

(i) Traditional insurance is group-based. Assuming that insurers use AI to make more accurate predictions by analysing more data, insurers will assign fewer people with the same attributes to the same risk pool. In other words, risk pools become smaller, but the underlying system is still group-based. With behaviour-based insurance, insurers can follow the concrete behaviour of *individuals*, and adapt insurance based on their real-time behaviour.[46]

(ii) In general, insurers focus on paying compensation for risks that unfold. Some small exceptions exist where insurers aim to prevent risks, such as an insurer giving a discount on home insurance if the consumer installs certified locks. Data-intensive underwriting still follows the traditional idea of insurance: compensating for claims based on group predictions. By contrast, with behaviour-based insurance, insurers focus more on preventing risks. For example, insurers could (at least in theory) prevent accidents by nudging risky drivers to drive safer.

In Table 1, we summarise the main differences between traditional insurance, data-intensive underwriting, and behaviour-based insurance.

**Table 1.** Comparison of traditional insurance, data-intensive underwriting, and behaviour-based insurance.

| | **Traditional insurance** | **Data-intensive underwriting** | **Behaviour-based insurance** |
|---|---|---|---|
| **Group/individual** | group characteristics | (more) group characteristics | individual behaviour |
| **Focus on compensation / prevention** | compensation | compensation | prevention |

elsevier.com/retrieve/pii/S0001457523003160>. Certain practices, such as credit scoring may also fall between the two trends. See e.g. The Geneva Association, *Responsible Use of Data in the Digital Age: Customer Expectations and Insurer Responses* (2022) 7–8 <www.genevaassociation.org/publication/new-technologies-and-data/responsible-use-data>.

[46]The discounts the insurer gives are not based on risk pooling, and this can cause tensions in the risk pooling of insurance. See Cevolini and Esposito (n 34) 130–31.

## 3. Non-discrimination and (other) unfair differentiation

In this paper, we refer to discrimination in the legal sense, which is narrower than in common English. Discrimination occurs only when an insurer differentiates between groups based on a legally protected characteristics, such as ethnicity or gender.[47] Such discrimination of protected groups can occur accidentally. Many non-discrimination statutes around the world focus on a limited list of protected characteristics. For example, the European Union's non-discrimination directives together prohibit discrimination for six protected characteristics: age; disability; gender; religion or belief; racial or ethnic origin; sexual orientation.[48] But some of the EU non-discrimination directives have a narrower scope, focusing on, for instance, the employment context.

For insurance, the EU non-discrimination directives only prohibit discrimination based on gender and ethnicity.[49] Member states of the EU must implement the directives into national law. Many member states extended the scope of the non-discrimination rules to other sectors, including the insurance sector.[50] We limit our discussion to gender and ethnicity, because those characteristics are specifically protected in EU-wide non-discrimination law.

Many forms of differentiation in insurance are not illegal under non-discrimination law, because they do not harm groups with protected characteristics. But differentiation between groups with other characteristics, that are not legally protected, may nevertheless feel unfair. We call that 'other unfair differentiation'.

[47]In general, ethnicity is the 'most protected' characteristic in the EUs non-discrimination directives. Watson writes: '[T]here appears to be a hierarchy of norms with gender and race discrimination given considerably more protection than discrimination on other grounds. This in itself leads to inequalities.' E Ellis and P Watson, *EU Anti-Discrimination Law* (Oxford University Press, 2012) 497. Indirect discrimination was introduced by the CJEU: 'the concept of indirect discrimination, and indispensable tool in the fight against discrimination, was developed by the Court in some of its earliest case law on sex equality and has now been written into all the equality directives' (497).

[48]Religion or belief, disability, age or sexual orientation: Council Directive 2000/78/EC of 27 November 2000 establishing a general framework for equal treatment in employment and occupation. Racial or ethnic origin: Council Directive 2000/43/EC Implementing the Principle of Equal Treatment Between Persons Irrespective of Racial or Ethnic Origin, 2000 OJ L 180/22. Gender: Council Directive 2004/113/EC Implementing the principle of Equal Treatment between Men and Women in the Access to and Supply of Goods and Services, 2004 OJ L 373/37. Directive 2006/54/EC of the European Parliament and of the Council on the implementation of the Principle of Equal Opportunities and Equal Treatment of Men and Women in Matters of Employment and Occupation (Recast), 2006 OJ L 204/23.

[49]Directive 2000/43/EC of 29 June 2000 implementing the principle of equal treatment between persons irrespective of racial or ethnic origin (Directive 2000/43/EC); Directive 2004/113/EC of 13 December 2004 implementing the principle of equal treatment between men and women in the access to and supply of goods and services (Directive 2004/113/EC).

[50]European Commission, Directorate General for Justice and Consumers, and European network of legal experts in gender equality and non discrimination, *A Comparative Analysis of Non-Discrimination Law in Europe 2022: The 27 EU Member States, Albania, Iceland, Liechtenstein, Montenegro, North Macedonia, Norway, Serbia, Turkey and the United Kingdom Compared* (Publications Office, 2023) 67 <https://data.europa.eu/doi/10.2838/428042>.

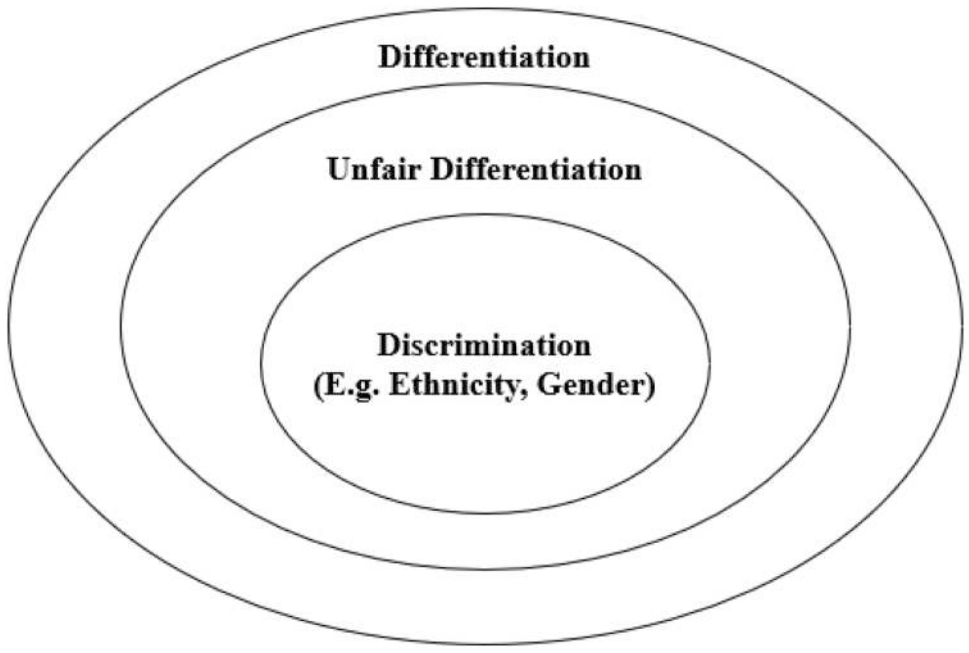

**Figure 1.** Discrimination and (unfair) differentiation. Discrimination is a type of differentiation that is considered unfair and can occur directly or indirectly. Unfair differentiation is a type of differentiation.

We see the relation between differentiation, other unfair differentiation, and discrimination as follows. With differentiation, we mean 'treatment of people or things in different ways'.[51] In that sense, discrimination is always a form of differentiation. Unfair differentiation is a form of differentiation. And discrimination in the legal sense is one type of unfair differentiation (Figure 1).[52]

## 4. Data-intensive underwriting

In this section, we discuss possible discriminatory effects of data-intensive underwriting (4.1) and possible unfair differentiation of data-intensive underwriting (4.2).

### *4.1. Discriminatory effects of data-intensive underwriting*

From a legal perspective, insurers may differentiate based on any characteristic, subject to some exceptions, namely legally protected characteristics such as gender and ethnicity.[53] Insurance is largely governed by contractual freedom: in principle, insurers choose with who they want to enter a contract, for what price, and under which conditions.[54] Non-discrimination law limits the contractual freedom of the insurer. To illustrate, because of non-discrimination law, an insurer is not allowed to refuse to deal with non-white people. While insurers are allowed to differentiate, they are not allowed to discriminate in the legal sense.[55] Insurers typically have more leeway in private insurance than in public insurances. For example, for health insurance, contractual freedom typically plays a smaller role. Depending on the country, the law may require health insurers to accept high-risk consumers, such as people with pre-existing diseases.[56]

[51] Cambridge Dictionary, 'Definition of Differentiation' <https://dictionary.cambridge.org/dictionary/english/differentiation> accessed 30 January 2025. The Geneva Association defines differentiation as 'any (lawful) differential treatment.' The Geneva Association (n 45) 9.

[52] We acknowledge that sometimes a decision may seem unfair to a consumer, but that does not make the decision as such unfair. Fairness and equality are explicit aims of non-discrimination law. Their aim is to give all individuals an equal and fair chance to access opportunities available in a society. See on the principle of non-discrimination: https://eur-lex.europa.eu/EN/legal-content/glossary/non-discrimination-the-principle-of.html.

[53] As noted, data protection law is outside the scope of this paper.

[54] Davey (n 21) 269.

In EU law, two forms of discrimination are prohibited: *direct* and *indirect* discrimination.[57] US law contains a similar distinction between *disparate treatment* and *disparate impact*. *Direct discrimination* means that organisations discriminate against people on the basis of a protected characteristic, such as ethnicity or gender. An extreme example of direct discrimination is, for example, the differentiation based on ethnicity in life insurance in the US before the 1950s. Insurers included ethnicity in their mortality tables, arguing that African Americans died, on average, sooner. From the 1950s, insurers stopped using ethnicity.[58]

In the words of the Racial Equality Directive, direct discrimination occurs 'where one person is treated less favourably than another is, has been or would be treated in a comparable situation on grounds of ethnic or ethnic origin'.[59] Direct discrimination is forbidden, except for a few specific and narrowly defined legal exceptions.[60] For the insurance sector, there are no exceptions in the non-discrimination directives that allow discriminating directly based on gender or ethnicity.

In the *Test-Achats* case from 2011, the Court of Justice of the European Union prohibited, roughly summarised, life insurers to discriminate directly based on gender. Until the judgment, many EU member states allowed gender-based price differentiation. EU non-discrimination law included a provision that allowed EU member states to adopt an exemption. The court declared that provision and the member states' exemptions invalid.[61] In sum, direct discrimination based on gender and ethnicity is prohibited for insurers.

---

[55]See Section 3, where we noted that we mean discrimination in the narrow, legal sense. See also Ronen Avraham, 'Discrimination and Insurance' in Kasper Lippert-Rasmussen (ed), *The Routledge Handbook of the Ethics of Discrimination* (Routledge, 2018).

[56]Public health insurance systems such as in, for example, The Netherlands and Germany are (partly) based on solidarity, not on risk calculations. See Johan van Manen, '15. Country Report: The Netherlands' in Wolf Sauter and others (eds), *The Law and Policy of Healthcare Financing* (Edward Elgar Publishing Limited, 2019) 360 <https://doi.org/10.4337/9781788115926>. See also Ulrich Von Ulmenstein and others, 'Limiting Medical Certainties? Funding Challenges for German and Comparable Public Healthcare Systems Due to AI Prediction and How to Address Them' (2022) 5 *Frontiers in Artificial Intelligence* 913093, 4 <www.frontiersin.org/articles/10.3389/frai.2022.913093/full>.

[57]For more details, see Frederik Zuiderveen Borgesius and others, *Non-Discrimination Law in Europe: A Primer for Non-Lawyers [PREPRINT]* (2024) <https://arxiv.org/abs/2404.08519>.

[58]Caley Dawn Horan, *Actuarial Age: Insurance and the Emergence of Neoliberalism in the Postwar United States* (University of Minnesota, 2011) 157–58, 186. Dan Bouk, *How Our Days Became Numbered* (University of Chicago Press, 2015) 32–35, 199–205.

[59]Article 2(2) of the Racial Equality Directive.

[60]F Zuiderveen Borgesius, 'Price Discrimination, Algorithmic Decision-Making, and European Non-Discrimination Law' (2020) 31 *European Business Law Review* 22, 408–09.

Even if insurers underwrite increasingly data-intensively, it may seem difficult to imagine an insurer discriminating directly based on ethnicity or gender. Yet perhaps unintentionally, such direct discrimination can occur. In 2013, in the Netherlands, an insurer excluded people living in caravans from home insurance. The non-discrimination authority decided that differentiating based on whether people were living in caravans constitutes direct discrimination, because all Roma were affected by the decision to exclude caravans.[62] The Dutch insurers maybe did not expect that the differentiation based on 'living in a caravan' amounts to direct ethnicity discrimination, because the insurers were using a proxy variable. While in general we do not think that data-intensive underwriting leads to more direct discrimination, it can still occur unintentionally or unexpectedly.[63]

More likely is that an insurer might discriminate indirectly. Indirect discrimination occurs when a practice is neutral at first glance but ends up discriminating against people with a certain protected characteristic. For example, suppose that a neighbourhood mostly includes people of a certain ethnicity. If an insurer charges higher prices in that neighbourhood, the insurer could accidentally discriminate against that ethnicity. Even if the insurer was unaware of the indirect discrimination, it is in principle prohibited.

However, if the insurer can invoke an 'objective justification', insurers do not indirectly discriminate. According to the non-discrimination directives, insurers can objectively justify indirect discrimination if they have 'a legitimate aim and [if] the means of achieving that aim are appropriate and necessary'.[64] An example of a legitimate aim in practice is if insurers try to lower their costs and reduce the administrative burden on their consumers.[65]

Judges seem to accept many aims as legitimate. But having a legitimate aim is not sufficient; the differentiation must be 'appropriate and necessary'. It depends on all circumstances of a case whether a practice is 'appropriate and necessary'. Generally speaking, the insurer must choose the least intrusive form of differentiation, and must ensure that a practice does not cause disproportionate disadvantages for protected groups.[66]

[61]Test-Achats [2011] CJEU C-236/09.

[62]NJCM, *Reaal Woonverzekeringen* <https://oordelen.mensenrechten.nl/oordeel/2018-14>. See also CJEU 12 December 2013, ECLI:EU:C:2013:823 (Hay) [44] <https://eur-lex.europa.eu/legal-content/EN/TXT/HTML/?uri=CELEX:62012CJ0267>.

[63]We note that in Europe, intentionality is not a requirement for direct discrimination.

[64]Article 2(2)(b) Directive 2000/43/EC of 29 June 2000 implementing the principle of equal treatment between persons irrespective of racial or ethnic origin (Directive 2000/43/EC).

[65]College voor de Rechten van de Mens, *Advies aan Dazure B.V. over premiedifferentiatie op basis van postcode bij de Finvita overlijdensrisicoverzekering* (2014) <https://publicaties.mensenrechten.nl/publicatie/29c613ac-efab-4d2a-a26c-fe1cb2556407>.

In data-intensive underwriting, insurers look for more data to further differentiate their risk pools. The threat of indirect discrimination may increase, especially if the insurers are not careful of which proxies they choose.

### 4.2. Other unfair differentiation of data-intensive underwriting

Which differentiation by insurers is fair and unfair? The line is difficult to draw. There is no uniformly accepted definition of fairness in the insurance context.[67] The fairness of differentiation depends on all the circumstances of a situation.[68] In the field of AI, many technical definitions of fairness exist, also with sometimes conflicting requirements. Different contexts may require balancing different notions of fairness.[69]

In this section, we analyse data-intensive underwriting according to four main factors that may indicate unfair differentiation:

(1) consumers have no influence on the price
(2) insurers use characteristics that seem irrelevant for the consumer
(3) insurers reinforce financial inequality
(4) consumers are excluded from insurance

We base the four factors by combining various strands of literature. We do not argue that each factor always leads to unfairness. And in some situations, it may be unclear which factor carries more weight.[70] An insurance differentiation can be, or feel, unfair for consumers, without any ill intent by the insurer.

#### 4.2.1. No influence on the price

In general, a risk assessment is fairer if a consumer can at least influence the price. It can feel unfair for consumers if they pay a higher price, and they cannot do anything about it.[71] From the perspective of a consumer, taking a risk can be either a gamble or a choice. In a chapter about luck and insurance, Dworkin distinguishes between two types of luck: 'brute luck' and 'option luck':

> Option luck is a matter of how deliberate and calculated gambles turn out – whether someone gains or loses through accepting an isolated risk he or she should have anticipated and might have declined. Brute luck is a matter of how risks fall out that are not in that sense deliberate gambles. If I buy a stock on the exchange that rises, then my option luck is good. If I am hit by a falling meteorite whose course could not have been predicted, then my bad luck is brute […].[72]

[66]ERG and Others [2010] CJEU C 379/08 & C 380/08, ECLI:EU:C:2010:127 [86].
[67]Avraham (n 55); Duncan Minty, 'Why Equality of Fairness Will Shape the Future of Insurance' (*Ethics and Insurance*, 3 March 2021) <www.ethicsandinsurance.info/equality-of-fairness/> accessed 30 January 2025.
[68]Frezal and Barry (n 35) 134. Jyri Liukko, 'Genetic Discrimination, Insurance, and Solidarity: An Analysis of the Argumentation for Fair Risk Classification' (2010) 29 *New Genetics and Society* 457, 471 <www.tandfonline.com/doi/full/10.1080/14636778.2010.528197>. Rick Swedloff, 'The New Regulatory Imperative for Insurance' (2020) 61 *Boston College Law Review* 2031.
[69]See e.g. Solon Barocas, Moritz Hardt and Arvind Narayanan, *Fairness and Machine Learning* (MIT Press, 2023) <https://fairmlbook.org/>. See also Christian Fröhlich and Robert C Williamson, 'Insights From Insurance for Fair Machine Learning', *The 2024 ACM Conference on Fairness, Accountability, and Transparency* (ACM, 2024) 415 <https://dl.acm.org/doi/10.1145/3630106.3658914>.
[70]Different types of insurance may also require a different weighing of these factors. For example, public insurances versus private insurances.

Data-intensive underwriting may lead to more situations in which insurers base the price on factors that the consumer cannot reasonably influence, leading to a more brute luck for consumers. After all, insurers (or their AI systems) may use more and more types of data to predict claim costs and to set prices. For example, in 2022, a Dutch consumer organisation found that some insurers charge different car insurance prices based on the consumer's house number. A consumer paid more if their house number included a letter: for instance, 11B or 39A.[73] Consumers may feel that they cannot easily influence their house number. Moving to another house to influence one's insurance price is hardly a serious option. A similar argument could be made for postal codes. Consumers can influence other characteristics more easily, such as the type of car they buy.

#### *4.2.2. Seemingly irrelevant characteristics*

A strength of machine learning systems is that they can find new correlations in datasets. A weakness is that such systems do not help insurers in understanding the relationships in the data and explaining them. Barocas et al. write that 'A major limitation of machine learning is that it only reveals correlations, but [people] often use its predictions as if they reveal causation'.[74]

When insurers underwrite using more and more data, they might find many correlations in datasets that look weird. For example, a correlation between the claim cost of car insurance and even house numbers, being born in a certain month, or 'people who spend more than 50% of their days on streets starting with the letter J'.[75]

[71]See Laurence Barry and Arthur Charpentier, 'Melting Contestation: Insurance Fairness and Machine Learning' (2023) 25 *Ethics and Information Technology* 49, 49 <https://link.springer.com/10.1007/s10676-023-09720-y>. This is also called 'luck egalitarianism'. See Fröhlich and Williamson (n 69) 411.

[72]Ronald Dworkin, *Sovereign Virtue. The Theory and Practice of Equality* (Harvard University Press, 2000) 73.

[73]Consumentenbond, 'Premies verzekeringen verschillen tot op huisnummer' ['Insurance Prices Differ by House Number'] <www.consumentenbond.nl/inboedelverzekering/verzekeringspremies-verschillen-tot-op-huisnummer> accessed 30 January 2025.

[74]Barocas, Hardt and Narayanan (n 69) 13.

[75]Cathy O'Neil, *Weapons of Math Destruction: How Big Data Increases Inequality and Threatens Democracy* (Penguin Books, 2016) 138.

A possible threat of data-intensive underwriting is that insurers could apply new, opaque relationships for predicting the claim cost, even if they do not (fully) understand the relationship behind the correlation.[76] Some insurance companies may use such predictions if they increase profit.

Other insurers may have less faith in machine learning: They may only apply a certain relationship when they understand the causality behind the relationship.[77] Even if an insurer understands the causal relationship, however, consumers may still be confronted with a price based on characteristics that are – from the consumer's point of view – irrelevant.

We give an example. Suppose that a car insurer finds a correlation between a specific colour of car and the claim cost. It turns out that the colour of the car matches with the colour of many houses in certain cities, and is therefore more difficult for people to see in specific light conditions. The insurer sees people with the specific car colour as a separate risk pool for car insurance, and charges a higher price to consumers who drive a car with that colour.

While the insurer found an explanation for the relationship that is causally true, consumers may still find the characteristic irrelevant (or illogical), because it is difficult to understand how a specific car colour might create more accidents in specific situations. When requesting insurance, a consumer typically enters their car colour, and then sees the new price, without further explanation. We draw the example in Figure 2.

In sum, we see two possible situations in which insurers use characteristics to calculate prices, while the characteristics seem strange or irrelevant for consumers. First, the insurer may not understand the causality behind a correlation, but still use the correlation to set prices. Second, the insurer understands the causal relationship, but the consumer still does not see the logic behind the relationship. As insurers collect and analyse more data for their underwriting, both types may occur.

### *4.2.3. Reinforcing financial inequalities*

Insurance practices could reinforce inequalities that already exist in society, such as the difference between rich and poor. Practices that reinforce financial inequality in society are controversial – some might say unfair. We give a real-life example. A Dutch insurer offered life insurance over the internet. The insurer wanted to charge higher prices to people likely to die earlier, and therefore charged higher prices to poorer people. In the Netherlands, income correlates strongly with life expectancy: on average, poorer people die years earlier than richer people.[78] Income also correlates with postal code. Therefore, the insurer charged higher life insurance prices in poorer postal codes. Besides the postal code, the insurer also considered the consumer's age, smoking behaviour, and health information for the insurance price. The insurer contacted the national non-discrimination regulator to ask whether the law allowed its practice.

[76]For more information about causality, see Charpentier (n 4) 57.
[77]See Sections 2.2 and 2.3.

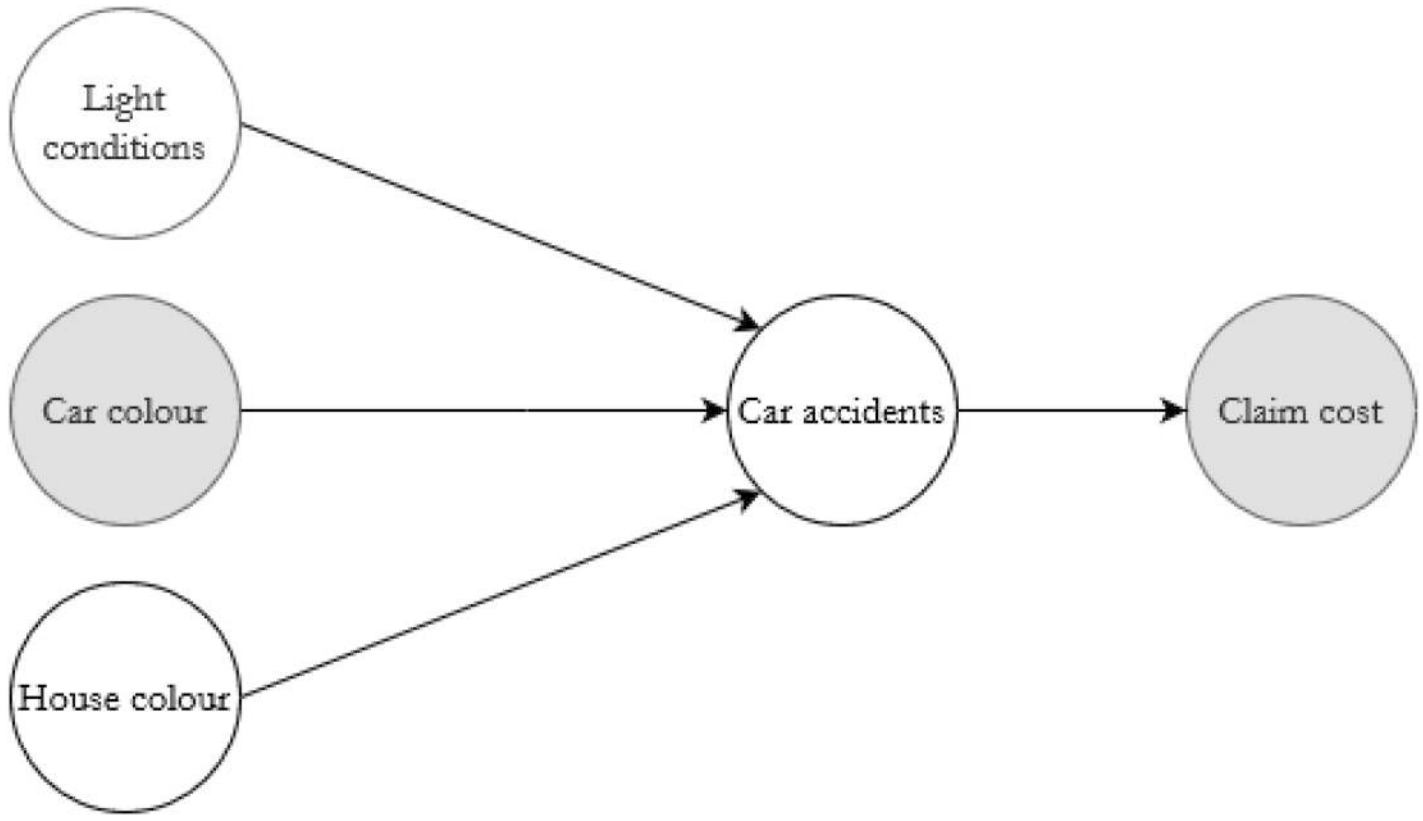

**Figure 2.** The arrows indicate the direction of the causal relationship. The grey nodes indicate what information about the specific consumer the insurer currently has: 'car colour' and 'claim cost'. The insurer predicts that light conditions, car colour and house colour cause more car accidents, which causes a higher claim cost. The consumer only sees that entering their car colour results in a higher claim cost.

The regulator consulted experts who confirmed that both postal code and income are a good predictor of life expectancy. The regulator ruled that the pricing was allowed because income or financial status are not protected grounds in Dutch non-discrimination law.[79] Therefore, such price differentiation is allowed by law. If the poor pay more, a practice reinforces financial inequality.[80] Many see it as unfair if a practice reinforces financial inequality.

Could data-intensive underwriting reinforce financial inequality in society? As discussed, an insurer could charge, on purpose, higher prices to poor people, because poor people live shorter on average. Similar situations may arise with data-intensive underwriting. First, an insurer's AI systems might find a correlation between being poor and filing more claims on average. The insurer might decide to charge higher prices to poor people. EU non-discrimination law does not prohibit such differentiation, unless the differentiation also disproportionally affects people of, for example, a certain ethnicity.[81]

[78]See, (in Dutch) Centraal Bureau voor de Statistiek [Dutch Central Bureau of Statistics], 'StatLine – Gezonde levensverwachting; inkomen en welvaart' (2022) <https://opendata.cbs.nl/#/CBS/nl/dataset/85445NED/table?defaultview&dl=747D6> accessed 31 January 2025.
[79]College voor de Rechten van de Mens (n 65) 16.
[80]The phrase 'the poor pay more' comes from a book, see David Caplovitz, *The Poor Pay More: Consumer Practices of Low-Income Families* (Free Press of Glencoe, 1963).

Second, data-intensive underwriting could also lead to more situations where insurers *unintentionally* differentiate based on income. Take the following hypothetical. An insurer's AI system finds a correlation that improves the prediction of expected claims. Let's say that the insurer finds a new group (risk pool), that is distinguished by a complicated set of characteristics.[82] The insurer does not fully understand the causality behind the AI prediction, but the prediction is accurate in practice. The insurer charges higher prices to that group. Unbeknownst to the insurer, that group is, on average, poor. Hence, an insurer might charge higher prices to a certain group, while the insurer does not realise that this harms poor people.[83]

We do not claim that the poor will always pay more because of data-intensive underwriting. There may be situations where data-intensive underwriting will lead to opposite: the rich paying more. Depending on the situation, data-intensive underwriting could reinforce or mitigate financial inequality.

#### *4.2.4. Exclusion*

Insurers could exclude consumers from insurance in several ways, intentionally or not. First, insurers can refuse to insure somebody. For instance, a car insurer can refuse to contract with high-risk drivers. Second, an insurer may charge certain high-risk groups higher prices. Suppose that an insurer divides the total population in increasingly smaller risk pools. At some point, the price could become so high for the riskiest consumers that the insurer excludes those consumers de facto, even if the insurer does not formally refuse to contract with them.[84] Third, an insurer could advertise that it specialises in products for a certain group. For example, the Dutch insurer *Promovendum* says in its advertising that it targets only the higher educated. In reality, the insurer does not refuse applications by lower educated consumers, especially for mandatory insurances.[85] However, due to targeted advertising, most lower educated consumers will not contact this particular insurer. Hence, de facto the insurer excludes certain groups.[86]

[81]Some national non-discrimination laws do prohibit discriminating on financial status, but the prohibition is not EU-wide. See, in Dutch: Ashley Terlouw, 'Klassisme: discriminatie op grond van sociale status' ['Classism: Discrimination Based on Social Status'] (2022) 4 Nederlands Tijdschrift voor de Mensenrechten <https://njcm.nl/wp-content/uploads/2023/01/1.-NTM-47-4_Terlouw.docx.pdf>. See also Sarah Ganty and Juan Carlos Benito Sanchez, *Expanding the List of Protected Grounds within Anti-Discrimination Law in the EU* (Equinet, 2021) <https://equineteurope.org/publications/expanding-the-list-of-protected-grounds-within-anti-discrimination-law-in-the-eu-an-equinet-report/>.

[82]AI systems may identify new groups in society that deserve protection. See e.g. Sandra Wachter, 'The Theory of Artificial Immutability: Protecting Algorithmic Groups under Anti-Discrimination Law' <https://arxiv.org/abs/2205.01166>.

[83]Financial status is not an explicit characteristic under EU non-discrimination law. See section 3.1.

[84]Avraham (n 55) 342.

[85]See, in Dutch: Promovendum, 'Hoger opgeleiden [Higher Educated]' (*Promovendum*) <www.promovendum.nl/hoger-opgeleiden> accessed 30 January 2025.

Data-intensive underwriting brings a threat that certain groups in society are excluded from insurance. Insurers may become better at predicting who is a very high-risk consumer. For very high-risk consumers, prices may become unaffordable. Suppose that a care insurer charges a very high price to the riskiest drivers (say 5% of the population). The insurance could become unaffordable for those drivers.

Many insurers realise that underwriting more data-intensively may lead to non-insurability of groups of consumers. To illustrate, the Dutch Association of Insurers developed a 'solidarity monitor' in response to the threat of non-insurability. The Association tries to assess whether groups are excluded by looking at, for example, the spread of the price between groups and the percentage of people who were denied insurance. The Association reports every year on their findings. So far, the Association did not find evidence that the threat materialised in the Netherlands.[87]

Some authors worry that more precise underwriting can undermine the solidarity that was traditionally part of insurance: risks would no longer be cross-subsidised between riskier and less risky people in society.[88] Whether it is unfair if people are excluded from insurance, or large price difference appear, depends on many factors. For instance, the importance of an insurance product in society is relevant. It makes a difference whether somebody is excluded from health insurance for themselves, or from health insurance for their pet. Another factor to consider is whether the consumer has an alternative. If one insurer excludes certain consumers, but they can go to plenty of competitors, the exclusion problem is mitigated. In sum, data-intensive underwriting could lead to certain groups being excluded from insurance.

### 4.3. Conclusion

In this section, we discussed the trend of data-intensive underwriting and analysed the trend based on non-discrimination law and fairness considerations. Insurers may (unintentionally) discriminate indirectly more often. For example, an insurer could use a newly found correlation to set prices.

[86]In 2014, a report by several Dutch governmental advisory organs named the insurance as a possible sign of explicit segregation. See in Dutch: Sociaal en Cultureel Planbureau (SCP) and Wetenschappelijke Raad voor het Regeringsbeleid (WRR), *Gescheiden Werelden? Een Verkenning van Sociaal-Culturele Tegenstellingen in Nederland [Separate Worlds? An Exploration of Socio-Cultural Divisions in the Netherlands]* (WRR, 2014) 24–25 <www.wrr.nl/publicaties/publicaties/2014/10/30/gescheiden-werelden-een-verkenning-van-sociaal-culturele-tegenstellingen-in-nederland>.

[87]Verbond van Verzekeraars, *Solidariteitsmonitor* [Solidarity Monitor] (2022) 7 <www.verzekeraars.nl/media/10617/solidariteitsmonitor-2022.pdf>. EIOPA (n 13).

[88]Frezal and Barry (n 35) 134.

Such a practice could harm certain ethnic groups, or other groups with legally protected characteristics, without the insurer realising it.

We found more possible negative effects of data-intensive underwriting. First, consumers may have less influence on the insurance price if insurers use new types of data for calculating prices. Second, insurers are more likely to choose characteristics that consumers find irrelevant, even if insurers find them relevant. Third, data-intensive underwriting may lead to more financial inequality in society. Fourth, insurers could exclude certain groups. Many of these effects could occur by mistake, rather than intentionally.[89] We repeat that we do not claim that all these effects will occur; we merely flag possible effects.

## 5. Behaviour-based insurance

Next, we discuss whether behaviour-based insurance may lead to more cases of discrimination (section 5.1) or other unfair differentiation (section 5.2).

### *5.1. Discriminatory effects of behaviour-based insurance*

Can behaviour-based insurance lead to indirect discrimination of legally protected groups? If insurers set the price based on the consumer's behaviour, discrimination seems unlikely at first glance. For example, driving behaviour is not a proxy for ethnicity or gender. Therefore, an insurer does not discriminate directly if it sets a price based on driving behaviour.

There is also no reason to assume indirect discrimination: driving behaviour is a neutral criterion, and there is no evidence to suggest that certain ethnicities are disproportionally affected if insurers set prices based on driving behaviour. And even if an insurer's practice affects mostly certain ethnicities, an insurer might objectively justify the indirect discrimination by stating that a consumer's driving behaviour is often the direct cause of an accident, that consumers can easily influence their driving behaviour, and that the behaviour is therefore relevant. If an insurer has an objective justification, the prima facie discrimination is not illegal. Nevertheless, in theory, ethnicity might correlate with driving behaviour. Hence, in theory, behaviour-based car insurance could lead to illegal discrimination.

Second, it is possible that, for instance, driving behaviour correlates with certain genders.[90] In Europe, insurers are not legally allowed to charge different car insurance prices directly based on a person's gender. However, if women generally drive more carefully than men, women will generally pay lower prices with behaviour-based car insurance. Such an indirect price difference seems justifiable. A man could drive more carefully, and thus receive the lower price too. A similar argument could be made for fitness trackers and discounts on insurance prices: irrespective of gender, men and women could walk more often. To sum up, behaviour-based insurance seems unlikely to lead to more cases of illegal discrimination.

[89]Minty argues that there should be more debate between insurers and consumers, to ensure that insurance is and remains fair. Duncan Minty, *Revolutionising Fairness to Enable Digital Insurance* (Institute and Faculty of Actuaries 2023) 11, 15–16 <https://actuaries.org.uk/media/z5rh5jhl/revolutionising-fairness-to-enable-digital-insurance.pdf>.

[90]See also Infantino (n 4) 626.

### 5.2. Other unfair differentiation of behaviour-based insurance

In this section, we discuss possible unfair differentiation as a result of behaviour-based insurance. We discuss the same factors as for data-intensive underwriting: no influence on the price (5.2.1), irrelevant characteristics (5.2.2), reinforcing financial inequality (5.2.3) and exclusion (5.2.4).

#### 5.2.1. No influence on the price

Will consumers see behaviour-based insurance as unfair because they cannot influence the price? That seems unlikely. For instance, consumers can influence the price, by driving more carefully or by walking more steps.[91] Consumers may therefore see behaviour-based insurance prices as fairer than the current insurance price, which are often based on characteristics that they have little influence on. On the other hand, consumers may not be able to control behaviour over a longer period. First, changing a habit can take a long time, often months. It is therefore questionable how easy changing behaviour truly is.[92] To answer the question of whether consumers can influence driving behaviour, input from psychology and other disciplines is needed.[93] All in all, consumers probably feel that they can influence the price with behaviour-based insurance.

#### 5.2.2. Seemingly irrelevant characteristics

Behaviour-based insurance is based on the idea that consumers can influence the price by behaving in a certain way. Therefore, it seems unlikely that consumers see the behaviour as an irrelevant characteristic. Consumers receive a discount if they have an active lifestyle (monitored by the insurer through a fitness tracker) or drive carefully (monitored through a device in the car). In these cases, the consumer probably sees the connection between their own behaviour and the odds that they file an insurance claim for accidents.[94]

[91] See section 4.2.1 for the distinction between brute and option luck.

[92] Stroobants and Van Schoubroeck (n 8) 79–80.

[93] See also Stroobants and Van Schoubroeck (n 8) 80. For a review of social scientific perspectives on behaviour-based insurance, see Maiju Tanninen, 'Contested Technology: Social Scientific Perspectives of Behaviour-Based Insurance' (2020) 7 *Big Data & Society* 205395172094253 <https://journals.sagepub.com/doi/10.1177/2053951720942536>.

Another question is whether, for example, the way people drive also affects the claims cost of that group: does more careful driving actually lower the claims cost for the insurer? A survey by Swiss Re from 2017 suggests that while safer drivers generally file less claims (−2% to −5%), their average claim cost per accident is higher (3%) than for drivers without behaviour-based insurance.[95] Nevertheless, we expect that most consumers think that their behaviour is relevant for the price of behaviour-based insurance.

### *5.2.3. Reinforcing financial inequality*

With behaviour-based insurance, an insurer monitors a consumer's behaviour, and not their financial status. Hence, at first glance, it seems unlikely that behaviour-based insurance can reinforce financial inequality. Nevertheless, some forms of behaviour-based insurance can benefit the rich more than the poor. In such cases, behaviour-based insurance could reinforce financial inequality.

To give an example for health trackers, suppose that many richer and better-educated people have jobs with more freedom, such as the possibility to work from home. For somebody who works at home, it is easier to go for a run or to the gym during the day than, for example, a factory worker. Somebody working from home can catch up on work at night. If a life insurer gives discounts to people with a more active lifestyle, on average, richer people may get more discounts. Such a situation would reinforce financial inequality.

However, the contrary effect is also possible: perhaps office workers have a less active lifestyle than less well-paid people who work in, for example, warehouses, or who deliver the mail. On average, poorer people could have more active lifestyles. Hence, an insurer who gives discounts to people with a more active lifestyle would not reinforce financial inequality.

Could behaviour-based car insurance reinforce financial inequality? At first glance, one's driving behaviour does not seem to correlate with financial status. However, it might be the case that driving style (as measured by a box in a car) correlates with financial status. For example, perhaps people with lower-educated and lower-income jobs work night shifts more often, and are on average more tired than people with high-income jobs. People who are more tired may drive less carefully. In this case, driving less carefully would correlate with being poorer. Thus, behaviour-based insurance could reinforce financial inequality.

[94]We can see exceptional cases where a fitness tracker is not relevant for life expectancy, such as people in a wheelchair who, because of their disability, cannot walk.

[95]Swiss Re (n 39) 24 & 29.

But the contrary effect also seems possible. Some well-paid jobs may make people extra tired and thus less careful drivers: think of partners in law firms who work too many hours. There is another unequal effect of behaviour-based insurance: richer people can afford to refuse it. Consumers with more money can forego a discount if they dislike the privacy interference of the insurer's tracking their behaviour. Such differences could reinforce inequality.

In sum, it is impossible to say, in general terms, whether behaviour-based insurance will reinforce or mitigate financial inequality. In some cases, behaviour-based insurance might reinforce inequality; in other cases it might mitigate inequality.

#### *5.2.4. Exclusion*

On the one hand, behaviour-based insurance could lead to more *inclusion*. As the European Insurance and Occupational Pensions Authority (EIOPA), an EU supervisory body, puts it,

> a better understanding of the risks in combination with risk-mitigation services can improve financial inclusion for some high-risk consumers who previously could not access affordable coverage. Examples include young drivers using telematics devices and patients with diabetes using health wearable devices.[96]

Indeed, if behaviour-based car insurance did not exist, perhaps many younger people could not afford car insurance at all.[97] In the United Kingdom, behaviour-based insurance seems necessary for many young drivers to afford car insurance: the prices are unaffordable otherwise.[98]

Behaviour-based insurance may also exclude certain groups. Insurers using behaviour-based insurance might charge such high prices to some people that the insurance becomes unaffordable for them.[99] On average, younger car drivers have more accidents.[100] Suppose that younger people are poorer than older people. Basing insurance prices on monitoring people's driving style may lead to higher, and possibly unaffordable, prices for younger people. This problem is not new, however: without monitoring driving behaviour, many car insurers already charge higher prices to younger drivers.

[96]European Insurance and Occupational Pensions Authority (EIOPA) (n 1) 25 & 10.

[97]Dutch Financial Authority (AFM), *The Personalisation of Prices and Conditions in the Insurance Sector. An Exploratory Study* (AFM 2021) 21 <www.afm.nl/en/sector/actueel/2021/juni/aandachtspunten-gepersonaliseerde-beprijzing>. Tzameret H Rubin, Tor Helge Aas and Jackie Williams, 'Big Data and Data Ownership Rights: The Case of Car Insurance' (2023) 13 *Journal of Information Technology Teaching Cases* 82, 82–83 <https://doi.org/10.1177/20438869221096859>.

[98]Dutch Financial Authority (AFM) (n 97) 21. Matthew Jenkin, '"I Was Quoted £2,000 for Annual Cover": Are Young People Being Priced out of Car Insurance? – Which? News' (*Which?*, 1 October 2023) <www.which.co.uk/news/article/are-young-people-being-priced-out-of-car-insurance-aSHpn4t4mz4k> accessed 30 January 2025. Rubin, Aas and Williams (n 97) 82–83.

[99]Barry and Charpentier, 'Melting Contestation' (n 71) 49.

[100]See European Commission, *Facts and Figures Young People* (European Road Safety Observatory. Brussels, European Commission, Directorate General for Transport, 2021) <https://road-safety.transport.ec.europa.eu/system/files/2022-01/F%26F_young_people_20211221.pdf>.

### 5.3. Conclusion and discussion

We conclude that generally, behaviour-based insurance seems unlikely to cause discrimination against certain ethnicities or genders. But it is possible that behaviour-based insurance could lead to unfair differentiation in some cases.

Consumers can often influence their short-term behaviour. Behaviour seems relevant for behaviour-based car and life insurance. Behaviour-based insurance could reinforce or mitigate financial inequality, depending on the situation. Finally, behaviour-based insurance could lead to more inclusion, but also exclusion: insurers could exclude some people by focusing on behaviour, because insurance prices become too high. But behaviour-based insurance might also enable other consumers to obtain insurance, for whom insurance is too expensive without the option.

Some commenters wonder whether behaviour-based insurance is a 'fair alternative' to traditional insurance.[101] Because behaviour-based insurance focuses on the individual's characteristics (influenceable behaviour) rather than group characteristics that the individual cannot influence, some say that behaviour-based insurance is *actuarially fairer*. Roughly speaking, actuarial fairness means that insurers treat similar risks in similar ways: the price that an individual pays corresponds to the actual risk. But actuarial fairness is not the only type of fairness: there are other aspects to consider.[102] Our analysis of other aspects of fairness (controllable behaviour, relevance, financial inequality, and exclusion) did not show that behaviour-based insurance is generally fairer than traditional insurance.

[101] Barry and Charpentier write, in their interpretation: 'The 2011 European Gender Directive, which prohibits the use of gender in pricing can be seen as […] an invitation from the judge to adopt algorithmic tools and behavioral data instead of broad classes.' Barry and Charpentier, 'Melting Contestation' (n 71) 49. See also Lisa Rebert and Ine Van Hoyweghen, 'The Right to Underwrite Gender: The Goods & Services Directive and the Politics of Insurance Pricing' (2015) 18 *Tijdschrift voor Genderstudies* 413 <www.aup-online.com/content/journals/10.5117/TVGN2015.4.REBE>. See also Laurence Barry and Arthur Charpentier, 'The Fairness of Machine Learning in Insurance: New Rags for an Old Man?' s 4 <http://arxiv.org/abs/2205.08112> accessed 30 January 2025.

[102] We will not summarize the whole discussion in the paper. For more details, see e.g. Saurabh Jha, 'Punishing the Lemon: The Ethics of Actuarial Fairness' (2012) 9 *Journal of the American College of Radiology* 887 <https://linkinghub.elsevier.com/retrieve/pii/S1546144012005418>; Fröhlich and Williamson (n 69) 411. Duncan Minty, 'Behavioural Fairness Is a Serious Risk to the Future of Insurance' (*Ethics and Insurance*, 29 April 2020) <www.ethicsandinsurance.info/behavioural-fairness/> accessed 30 January 2025. Fröhlich and Williamson (n 69) 412. Liz McFall, 'Personalizing Solidarity? The Role of Self-Tracking in Health Insurance Pricing' (2019) 48 *Economy and Society* 52 <www.tandfonline.com/doi/full/10.1080/03085147.2019.1570707>. The Geneva Association, *Research Brief. Promoting Responsible Artificial Intelligence in Insurance* (The Geneva Association – International Association for the Study of Insurance Economics) <www.genevaassociation.org/sites/default/files/ai_in_insurance_web_0.pdf>.

## 6. Research agenda

Based on the findings in our paper, we suggest possible directions for future research into discrimination and unfair differentiation by AI in insurance. We see many possibilities for exciting interdisciplinary research, for instance for actuaries, economists, ethicists, legal scholars, sociologists, and computer scientists.

For analytical purposes, we distinguished data-intensive underwriting and behaviour-based insurance. But the two practices cannot be neatly separated, as the practices can partly overlap. Future research could focus on borderline cases. For instance, insurers could collect data about individuals through behaviour-based insurance, and aggregate those data to build group profiles for data-intensive underwriting. Do insurers do this, and should that influence the normative analysis?

We used four factors to assess unfair differentiation, but we do not claim that they form a complete framework. We think of the factors as a starting point for further discussion and research. Future work could extend our list of factors to build a more comprehensive fairness framework for insurance.

Many of the effects we identified in the paper raise normative questions that could be researched more in-depth. For example: when is it acceptable if insurance practices reinforce financial inequality? Is it fair if people who are born clumsy pay higher insurance prices in behaviour-based car insurance? Is it fair if people bear the costs of traits or behaviour that they cannot control? Is behaviour-based insurance fairer than traditional insurance or data-intensive underwriting? Certain types of insurance call for a specific normative analysis, such as health insurance, where the right to healthcare should play a role. This paper did not focus on the privacy and surveillance aspects of behaviour-based insurance, which also deserve more attention. Some issues that arise in the context of AI may also need more discussion and debate where AI does not play a role, such as the fairness of excluding consumers from insurance.

Many of the topics in our paper deserve more empirical research. For example, will both trends identified in the paper continue? In the paper, we highlighted *possible* discriminatory and other unfair effects. Which effects will materialise in practice, if any? And to what extent is behaviour-based insurance based on empirical evidence? Does monitoring driving behaviour actually predict accidents, and are the AI models insurers use to assess driving behaviour correct, or a form of 'snake oil' AI?[103] Further

[103] According to Naranyanan, most snake oil AI is 'concentrated in predictive AI'. Arvind Narayanan, 'How to Recognize AI Snake Oil' [2019] Arthur Miller Lecture on Science and Ethics <www.cs.princeton.edu/~arvindn/talks/MIT-STS-AI-snakeoil.pdf>. Frederike Kaltheuner (ed), *Fake AI* (Meatspace Press, 2021) <https://fakeaibook.com/>. Arvind Narayanan and Sayash Kapoor, *AI Snake Oil. What Artificial*

research could investigate whether behaviour-based insurance actually influences consumer behaviour. And what do consumers and insurers think? Consumer surveys and interviews could provide insights about their opinions about AI in insurance. How do insurers use AI in practice? Interviews with insurers could provide valuable insights.

This paper focuses on risk-based pricing, which plays a large role in insurance pricing. But insurers may also adapt prices based on other factors, such as the expected willingness to pay of the consumer or the consumer's credit rating. Insurers could use AI to predict the maximum price a consumer is willing to pay. An insurer could, for instance, charge higher prices to consumers who pay little attention to prices. Such non-risk-based pricing is often called price discrimination in economic literature.[104] Future work could analyse the interplay between the effects of risk-based pricing and price discrimination.

While we discussed on possible effects, we did not discuss how insurers and policymakers could avoid or mitigate the effects. Open legal questions include: to what extent can existing law, such as non-discrimination law, consumer protection law, data protection law, and the AI Act, protect consumers and society against unfair effects of AI in insurance? If legal protection leaves gaps, should the law be improved, and how?

Finally, there are exciting possibilities for AI and computer science research. Could insurers make or use AI systems that are more transparent and explainable? Could sector-specific fairness and non-discrimination norms be built into AI systems?

Some questions that arise in the insurance sector are important in other sectors too. Is it fair if banks refuse to give a consumer credit because of seemingly irrelevant characteristics, because the bank's AI systems found some correlations? When is it acceptable if the use of AI increases financial inequality in society?

## 7. Summary and conclusion

In this paper, we discussed potential effects related to discrimination and unfair differentiation of insurers' use of AI. The paper explored which effects may occur if insurers (i) analyse more and new types of data (data-intensive underwriting), and (ii) adapt the price based on the consumer's real-time behaviour (behaviour-based insurance). Both trends are enabled by AI. We distinguished discrimination (which is prohibited by non-discrimination law) from other unfair differentiation. We defined

*Intelligence Can Do, What It Can't, and How to Tell the Difference* (Princeton University Press, 2024) <https://press.princeton.edu/books/hardcover/9780691249131/ai-snake-oil>.

[104] F Zuiderveen Borgesius and J Poort, 'Online Price Differentiation and EU Data Privacy Law' (2017) 40 *Journal of Consumer Policy*.

**Table 2.** Summary of conclusions.

| | Effect | Data-intensive underwriting | Behaviour-based insurance |
|---|---|---|---|
| **Discrimination** | *Direct or indirect discrimination* | Indirect discrimination could happen more often | Unlikely |
| **Unfair Differentiation** | *No influence on the price* | Could happen more often (more characteristics with less influence) | Unlikely (consumers can influence their behaviour) |
| | *Seemingly irrelevant characteristics* | Could happen more often (AI can find unexpected correlations) | Unlikely (behaviour seems relevant to consumer) |
| | *Reinforcing financial inequality* | Insurers could (unintentionally) reinforce financial inequality | Could reinforce but also mitigate financial inequality |
| | *Exclusion* | Could happen more often | Could happen more often |

discrimination as a form of legally forbidden differentiation consisting of direct and indirect discrimination. Unfair differentiation is a type of differentiation that seems unfair to the consumer.

We described many possible discrimination- or unfair differentiation-related effects of data-intensive underwriting and behaviour-based insurance. Table 2 gives a rough summary of the possible effects. We cannot conclude that data-intensive underwriting is fairer than behaviour-based insurance, or vice versa. Many aspects of data-intensive underwriting and behaviour-based insurance are unclear and deserve more research; we provided an interdisciplinary research agenda. One thing is clear, however: public debate is needed to decide which insurance practices we accept in our societies.

## Acknowledgements

The authors are especially grateful to Bart van der Sloot (Tilburg University), Paola Lopez (University of Vienna) and Balázs Bodó (University of Amsterdam) for their insightful comments during the PLSC-style session of Digital Legal Talks 2023. We also thank Jos Schaffers (Dutch Association of Insurers) and Maaike Harbers (Rotterdam University of Applied Sciences). We also thank the participants of the workshop ‘Insurance, algorithmic decision-making, and discrimination’ at iHub, Radboud University (2022).

## Disclosure statement

No potential conflict of interest was reported by the author(s).

## Notes on contributors

*Marvin van Bekkum* is a PhD candidate at Radboud University.

*Frederik Zuiderveen Borgesius* is a Professor of ICT and Law at Radboud University.

*Tom Heskes* is a Professor of Data Science at Radboud University.

## ORCID

*Marvin van Bekkum* http://orcid.org/0000-0002-8031-0134
*Frederik Zuiderveen Borgesius* http://orcid.org/0000-0001-5803-827X
*Tom Heskes* http://orcid.org/0000-0002-3398-5235